\documentclass[prl,twocolumn]{revtex4}
\usepackage[dvips]{graphicx}
\usepackage{latexsym}
\begin{document}
\renewcommand{\ni}{{\noindent}}
\newcommand{\dprime}{{\prime\prime}}
\newcommand{\be}{\begin{equation}}
\newcommand{\ee}{\end{equation}}
\newcommand{\bea}{\begin{eqnarray}} 
\newcommand{\eea}{\end{eqnarray}}
\newcommand{\la}{\langle}
\newcommand{\ra}{\rangle} 
\newcommand{\dg}{\dagger}
\newcommand\lbs{\left[}
\newcommand\rbs{\right]}
\newcommand\lbr{\left(}
\newcommand\rbr{\right)}
\newcommand\f{\frac}
\newcommand\e{\epsilon}
\newcommand\upa{\uparrow}
\newcommand\downa{\downarrow}
\newcommand\br{{\bf r}}
\newcommand\brp{{\bf r}^\prime}
\newcommand\bR{{\bf R}}
\newcommand\bk{{\bf k}}
\newcommand\bkp{{\bf k}^\prime}
\newcommand\cP{{\cal P}}
\title{Particle-Hole Asymmetry in Doped Mott Insulators:
Implications for Tunneling and Photoemission Spectroscopies}
\author{Mohit Randeria$^{(1,2)}$, Rajdeep Sensarma$^{(2)}$, Nandini Trivedi$^{(1,2)}$, Fu-Chun Zhang$^{(3)}$}
\affiliation{
(1) Physics Department, The Ohio State University, Columbus, OH 43210 \\
(2) Tata Institute of Fundamental Research, Mumbai 400 005, India \\
(3) Physics Department, University of Hong Kong, Hong Kong \\
{\ \ } Physics Department, University of Cincinnati, Cincinati, OH 45221}
\vspace{0.2cm}
\begin{abstract}
\vspace{0.3cm} 
In a system with strong local repulsive interactions it should be
more difficult to add an electron than to extract one. 
We make this idea precise by deriving various exact sum rules for 
the one-particle spectral function independent of the details 
of the Hamiltonian describing the system and of the nature of the ground state. 
We extend these results using a variational ansatz for the superconducting
ground state and low lying excitations. Our results shed light on the striking 
asymmetry in the tunneling spectra of high Tc superconductors and should 
also be useful in estimating the local doping variations in inhomogeneous
materials. 
\vspace{0.1cm}
\typeout{polish abstract}.
\end{abstract} 
\maketitle

With the discovery of high temperature superconductivity in the cuprates there has been
enormous interest in the properties of doped Mott insulators. One-particle spectroscopies
like angle resolved photoemission (ARPES) \cite{shen,campuzano} 
and scanning tunneling microscopy (STM) \cite{fischer,davis} have played a major
role in our understanding of these strongly correlated materials. In this paper we examine in detail
sum rule constraints on the single-electron spectral function, focusing in particular on the
{\em striking asymmetry between occupied and unoccupied spectral weights} in lightly doped Mott insulators
{\em and its doping dependence}.
Anderson \cite{anderson04} has recently emphasized that such an asymmetry 
is very unusual in conventional metallic systems, especially on low energy scales,
and has suggested that it may be a characteristic signature of ``projection'' into a low-energy subspace 
where strong local Coulomb repulsion makes double occupancy at a site energetically prohibitive.

In this letter, we first make this idea precise by deriving several
exact sum rules for the $T=0$ spectral function  for both the occupied and
unoccupied spectral weights in the low-energy subspace. These results are very general and 
do not depend either upon the details of the Hamiltonian or on any assumptions
about the nature of the ground state or low-lying excitations. We show that
these general results can be useful in several ways two ways. First, they help
in quantifying the particle-hole asymmetry in tunneling spectroscopy: 
there is much more weight on the negative bias (occupied) side
than on the positive bias (unoccupied) side \cite{davis1}.
Second, our results can be used in present day STM experiments to estimate the local 
doping variations in inhomogeneous systems \cite{davis2}.

We next separate out the coherent quasiparticle (QP) and the incoherent parts of the 
spectral function and predict their variation with hole doping in 
the d-wave superconducting state obtained upon doping a Mott insulator
\cite{anderson87,paramekanti,rvb_review}. 
This second set of results are variational in nature and require us to 
make assumptions about the ground state and low-lying QP excitations, and also to make the
Gutzwiller approximation \cite{rvb_review,zhang88} to obtain analytical results.
These results are testable in photoemission and inverse-photoemission experiments.

\medskip
\noindent{\bf Exact Sum Rules}

\medskip
Consider a system of electrons described by the Hamiltonian
$H = K + U\sum_i n_{i\upa} n_{j\downa}$
where $K$ is the Kinetic energy operator which can be an arbitrary tight binding Hamiltonian
with terms of order $t$ (nearest neighbor hopping). We can further add to $H$
a random one-body potential which could make the system inhomogeneous, 
as well as other potential energy terms such as longer range Coulomb interactions. 
We will always work in the limit where $U$ is much larger  than all other energy scales.
 
The one-electron spectral function is defined by
$A(\br,\brp;\omega) = {{-1}\over \pi}\ {\rm Im} G(\br,\brp; \omega + i0^+)$
where $G$ is the Green's function. We work in real space for two reasons:
first, the no-double occupancy constraint is best written in this basis, and 
second, this allows us to describe spatially inhomogeneous systems which is important
to discuss STM experiments on the cuprates.
We use the $T=0$ spectral representation
\bea
A(\br,\brp;\omega) = \sum_m \left[ 
   \langle 0 \vert c^\dagger_{\brp \sigma} \vert m \rangle 
   \langle m \vert c_{\br \sigma} \vert 0 \rangle \delta(\omega + E_m - E_0) \right] 
 \nonumber \\
 \mbox{~~~~~~~~} + \sum_m \left[ \langle 0 \vert c_{\br \sigma} \vert m \rangle 
   \langle m \vert c_{\brp \sigma}^\dagger \vert 0 \rangle 
   \delta(\omega - E_m + E_0) \right]
\label{akw}
\eea
where $\vert m \rangle$'s are exact many-body eigenstates with energy $E_m$
with $m=0$ the ground state, and $\omega$ is measured with respect to the
chemical potential.

The large $U$ suppresses double-occupancy at each site, and its effects on the
ground state and low-lying excitations are best described using the
the projection operator
$\cP = \prod_r \left( 1 - n_{r\uparrow}n_{r\downarrow} \right)$.
We then make the well-known unitary transformation \cite{unitary_transformation}
$\exp(-iS)$, such that $\exp(iS) H \exp(-iS)$ has no matrix elements connecting states
which differ in their double-occupancy, to any given order in $t/U$.
To leading order $iS = - (1/U) \sum_{\br,\brp,\sigma} t_{\br\brp}
\left( n_{\br\overline{\sigma}}c^\dagger_{\br\sigma}c_{\brp\sigma}h_{\brp\overline{\sigma}} 
- {\rm h.c.}\right) + {\cal O}(t/U)^2$, where $h_{\br\sigma} = 1 - n_{\br,\sigma}$.
It is useful to incorporate this unitary transformation on the states, 
which is equivalent to transforming {\em all} operators.
It then follows that {\em all the low-energy states}, i.e., those in the so-called
``lower Hubbard band" (LHB), are of the form $\exp(-iS) \cP \vert \Phi_m \rangle$
where $\vert \Phi_m \rangle$'s are {\em un}projected states. This characterization
of LHB states will be crucial below.

We now derive various exact sum rules without making any assumptions about the
nature of the ground state or low-lying excitations. Some of these are
very well known and shown only for completeness because we need to reference them 
later. From eq.~(\ref{akw}) it is trivial to see the total spectral weight
$\int_{-\infty}^{+\infty} d\omega A(\br,\brp;\omega) = 1$,
while the occupied spectral weight \cite{footnote1}
\begin{equation}
\int_{-\infty}^{0} d\omega A(\br,\brp;\omega) = 
\langle 0 \vert c^\dagger_{\brp \sigma}c_{\br \sigma} \vert 0 \rangle. 
\label{occupied}
\end{equation}

Without assuming translational invariance, 
the local density of states (LDOS) probed by STM experiments
$N(\br;\omega) = 2A(\br,\br;\omega)$, with the factor of two coming from
spin, is given by
\begin{equation}
\int_{-\infty}^{0} d\omega N(\br;\omega) = n(\br) = 1 - x(\br)
\label{occupied_STM}.
\end{equation}
Here $n(\br)$ is the local electron density and $x(\br)$ the local hole doping,
which for a translationally invariant system would be $\br$-independent.
This result simply says that there are $(1 - x)$ occupied sites (per unit volume) 
from which one can {\em remove} an electron.

For a translationally invariant system, we can Fourier transform 
(\ref{occupied}) from $(\br - \brp)$ to $\bk$ and obtain the well known result
$\int_{-\infty}^{0} d\omega A(\bk,\omega) = n(\bk)$ which has proved
useful in analyzing ARPES data \cite{randeria95}. 
Summing over all $\bk$'s and both spins one obtains 
$\int_{-\infty}^{0} d\omega N(\omega) = 1 - x$, which is
(\ref{occupied_STM}) for a uniform system.

Next we turn to sum rule constraints on the {\em unoccupied} side, which is the positive
bias side in tunneling or that probed by inverse photoemission.
It is trivial to derive sum rules for energy integration from $0$ (chemical potential)
to $\infty$ by subtracting the occupied spectral weights (\ref{occupied}) or
(\ref{occupied_STM}) from the total spectral weight of unity. But a much more
physically meaningful result is obtained by focusing only on the {\em low-energy 
states} in the ``lower Hubbard band'' (LHB) by integrating over 
$0 \le \omega \le \Omega_L$, where the upper cut-off $\Omega_L$
satisfies $t \ll \Omega_L \ll U$.  This is implemented
by restricting the sum over intermediate states in (\ref{akw}) to
LHB states $\vert m \rangle = \exp(-iS) P \vert \Phi_m \rangle$, as discussed above.

We thus write the integrated low energy spectral weight on the unoccupied side as
$\int_{0}^{\Omega_L} d\omega A(\br,\brp;\omega) =
\sum_m \langle \Phi_0 \vert \cP \tilde{c}_{\br\sigma} \cP \vert \Phi_m \rangle
\langle \Phi_m \vert \cP \tilde{c}^{\dagger}_{\brp\sigma} \cP \vert \Phi_0 \rangle$.
Here we have found it convenient to move the unitary transformation back
onto the operators and have introduced the notation: 
$\tilde{c}_{r\sigma} = \exp(iS) c_{r\sigma} \exp(-iS)$
and $\tilde{c}^\dagger_{\br\sigma} = \exp(iS) c^\dagger_{\br\sigma} \exp(-iS)$.
We now use $\sum_m \vert \Phi_m \rangle \langle \Phi_m \vert = 1$, since
the $\vert \Phi_m \rangle$'s are {\em un}projected states,
to obtain
\be
\int_{0}^{\Omega_L} d\omega A(\br,\brp;\omega) = 
\langle \Phi_0 \vert \cP \tilde{c}_{\br \sigma} \cP \tilde{c}^{\dagger}_{\brp\sigma} \cP \vert \Phi_0 \rangle.
\label{unocc_sumrule}
\ee
To simplify this, we must calculate $\cP \tilde{c}^{\dagger} \cP$. 
To order $t/U$ we find:
$\cP \tilde{c}^{\dagger}_{\br\sigma} \cP = h_{\br\overline{\sigma}}c^\dagger_{\br\sigma}\cP +
{1 \over U} \sum_{\br,\bR,\sigma'}t_{\br\bR} h_{\bR\overline{\sigma}'}
c^\dagger_{\bR\sigma'}c_{\br\sigma'}n_{\br\overline{\sigma}}c^\dagger_{\br\sigma}\cP$
and $\tilde{c}_{\br\sigma}$ is the hermitian conjugate, where $h_{\br\sigma} = 1 - n_{\br,\sigma}$ and
$\overline{\sigma} = -\sigma$. 

We thus obtain the sum rule for low energy spectral weight on the {\em unoccupied} side:
\be
\int_{0}^{\Omega_L} d\omega N(\br;\omega) = 2x(\br) + 2 \left\vert \langle K(\br) \rangle \right\vert / U
\label{unoccupied_STM}
\ee
where $\langle K(\br) \rangle = 
\langle \Phi_0 \vert \cP \sum_{\bR,\sigma}t_{\bR\br}\left(c^\dagger_{\bR\sigma}c_{\br\sigma} + {\rm h.c.}
\right) \cP \vert \Phi_0 \rangle$. 
The first term in (\ref{unoccupied_STM}) simply says that one can inject an electron into any of the 
$x$ empty sites, with the factor of two for spin degeneracy. The second term gives an order $(xt/U)$ 
correction since the injected electron can create a temporary double occupancy and then hop off to a 
neighboring empty site. We note that, in contrast to this, the corresponding result (\ref{occupied_STM})
to extract an electron is {\em exact} to all orders in $t/U$. 

For a translationally invariant system, we may rewrite the above result as
\be
\sum_{\bk}\int_{0}^{\Omega_L} d\omega A(\bk,\omega) = x + \left\vert \langle K \rangle \right\vert / U. 
\label{unoccupied_kspace}
\ee 
There is another simple result that can be obtained in the translationally invariant case.
First we simplify the right hand side of (\ref{unocc_sumrule})
using the lowest order expressions for $\cP \tilde{c}^\dagger \cP$ and $\cP \tilde{c} \cP$.
A straightforward calculation then shows that
$\int_{0}^{\Omega_L} d\omega A(\br,\brp;\omega) = (1 + x)/2 - 
\left\langle c^{\dagger}_{\brp\sigma} c_{\br\sigma} \right\rangle + {\cal{O}}(t/U)$.
Fourier transforming to $\bk$-space and using standard expressions of $n(\bk)$, we find that the
the total low-energy spectral weight is 
\begin{equation}
\int_{-\infty}^{\Omega_L} d\omega A(\bk,\omega) = {{1 + x}\over 2} + {\cal{O}}(t/U)
\label{total_low}
\end{equation}
{\em for each} $\bk$. Note that the deficit from unity comes from spectral weight in the ``upper 
Hubbard band'' which lies above $\Omega_L$. 

\medskip
\noindent{\bf Variational and Gutzwiller approximation results}
\medskip

We emphasize that no approximations were made to obtain the 
above results, and we also made no assumptions about the nature
of the ground state or low-lying excitations.

We now turn to translationally invariant systems and our goal is to
obtain more detailed information about the spectral function: to decompose 
it into its coherent and incoherent pieces and determine their doping dependence.
We take the (variational) ground state to be a projected d-wave 
BCS state $\vert 0 \rangle = \exp(iS) \cP \vert {\rm dBCS} \rangle$
which has given much insight into the phenomenology of the superconducting state of 
the high Tc cuprates \cite{anderson87,paramekanti}.
Further the (variational) quasiparticle (QP) excitations \cite{zhang88} above this
ground state are described by
$\vert k \sigma \rangle = \exp(iS) \cP \gamma^\dagger_{k \sigma} \vert {\rm dBCS} \rangle$,
where $\gamma^\dagger$ is the standard Bogoliubov QP operator.
The QP's lead to the {\em coherent} part of $A(\bk,\omega)$, i.e.,
delta-functions in $\omega$ at T=0.
Finally we make the Gutzwiller approximation (GA) \cite{zhang88,rvb_review}
which greatly simplifies calculations of matrix elements in this strongly 
interacting system and gives answers which are in good agreement with 
exact Monte Carlo results. 

Making a Gutzwiller approximation (GA) for the QP matrix elements in (\ref{akw})
we obtain \cite{rajdeep}
\bea
A(\bk,\omega) = Z(\bk) u_\bk^2 \delta(\omega - E_\bk) \mbox{~~~~~~~~~~~~}  \nonumber \\
\mbox{~~~~~~~~~} + Z(\bk) v_\bk^2 \delta(\omega + E_\bk) + A_{inc}(\bk,\omega)
\label{akw-ga}
\eea
where $u_\bk, v_\bk$ and $E_\bk$ is standard BCS notation. 
The first two terms in (\ref{akw-ga})are the coherent QP pieces 
with the same structure as in BCS theory, except that 
their spectral weight is suppressed by 
\bea
Z(\bk) = {2x \over {1 + x}} + {8x \over {U(1 + x)^2}} \sum_{\bkp} \epsilon_{\bkp} v^2_{\bkp}
\nonumber \\
\mbox{~~~~~~} + {4x \over {U(1 + x)}} \epsilon_{\bk} \sum_{\bkp} v^2_{\bkp} 
\label{z_factor}
\eea
where $\epsilon_{\bk}$ is the dispersion corresponding to the
bare kinetic energy $K$ in the Hamiltonian. We note that, as
emphasized in \cite{paramekanti}, $Z$ vanishes as
one goes to the insulating state at $x=0$, and, in fact, the GA result
(\ref{z_factor}) is in excellent quantitative agreement with the variational
Monte Carlo results of Paramekanti et al. \cite{paramekanti}.
The sum over all states other than single QP's in (\ref{akw}) leads to the
{\em incoherent} part of the spectral function denoted by
$A_{inc}(k,\omega)$. Although we cannot calculate its explicit form with the
minimal set of assumptions we have made, its existence is necessarily demanded 
by exact sum rules, as shown below, which also put constraints on $A_{inc}$.

Our use of the GA to calculate matrix elements goes beyond
previous applications of this approach, which have been by-and-large restricted to
ground state expectation values (see, however, the work of Laughlin \cite{laughlin} who
uses an approximation scheme closely related to GA for spectral functions). 
A non-trivial consistency check is provided by $n(\bk)$ calculated within GA, which involves only an 
equal-time ground state correlation and does not depend on any assumptions 
about QP excited states. We find $n(\bk) = Z(\bk) v_{\bk}^2 + n_{\rm smooth}(\bk)$ 
where $n_{\rm smooth}(\bk) = (1 - x)^2/2(1+x) + {\cal O}(t/U)$ is a smooth function
of $\bk$ in the entire Brillouin zone. We omit the details of the $(t/U)$ corrections 
here since they involve rather long expressions \cite{rajdeep}. We note that $n(\bk)$
implies that there is a jump discontinuity along the zone diagonal
whose magnitude is given precisely by (\ref{z_factor}) including the $(t/U)$ corrections.  

We now turn to sum rule constraints on $A_{inc}$ restricting ourselves, for the
most part, to leading order results in $t/U$; the next order corrections will be 
presented elsewhere \cite{rajdeep}. 
We begin by integrating (\ref{akw-ga}) from $-\infty$ to $0$ and comparing 
with the GA result for $n(\bk)$. We thus find that for each $\bk$
\begin{equation}
\int_{-\infty}^0 d\omega A_{inc}(\bk,\omega) = {(1-x)^2 \over 2(1+x)} + {\cal O}(t/U).
\label{inc-occ}
\end{equation}
Thus for each $\bk$ there is non-zero incoherent spectral weight 
for $\omega < 0$, whose strength relative to the coherent weight $Z(\bk)v_{\bk}^2$
on the occupied side, increases with underdoping (decreasing $x$).

To find the incoherent spectral weight on the unoccupied side
we substitute the GA spectral function (\ref{akw-ga}) in the total
low-energy spectral weight sum rule (\ref{total_low}), leading to
$\int_{-\infty}^{\Omega_L} d\omega A_{inc}(\bk,\omega) = (1-x)^2/[2(1+x)] + {\cal O}(t/U)$
for each $\bk$. This together with (\ref{inc-occ}) 
implies that $\int_{0}^{\Omega_L} d\omega A_{inc}(\bk,\omega) = {\cal O}(t/U)$.
Given the non-negativity of spectral weight, we find that 
\be
A_{inc}(\bk,\omega > 0) = {\cal O}(t/U) \ll 1.
\label{incoherent_unocc}
\ee
The {\em vanishing} of $A_{inc}$ for $\omega > 0$ to zeroth order in $(t/U)$ is at first sight
quite surprising. Although there is very little spectral weight on the
unoccupied side, as seen from (\ref{unoccupied_kspace}), whatever there is, is
entirely dominated by the coherent piece when $U \gg t$.

To gain more insight into this striking result we derive it in a completely different
fashion, which also shows that it is {\em not} an artifact of the Gutzwiller approximation.
To zeroth order in $(t/U)$ we can set $\exp(iS) = 1$, and using the identity
$\cP c^\dagger_{\br,\sigma} \cP = \cP c^\dagger_{\br,\sigma}$, we find that
$\cP c^\dagger_{\bk,\sigma} \vert 0 \rangle = \cP c^\dagger_{\bk,\sigma} \vert {\rm dBCS} \rangle
= {\rm constant} \times P \gamma^\dagger_{\bk,\sigma} \vert dBCS \rangle =
{\rm constant} \times \vert \bk,\sigma \rangle $.
We thus find that the projected creation operator acting on the ground state
gives precisely the coherent QP state. Thus there is no incoherent
weight in the electron creation (i.e., unoccupied) sector at least to
zeroth order in $(t/U)$. 
 
Finally, we can determine the explicit form of the order $t/U$
incoherent unoccupied spectral weight, which follows from use of eqs.~(\ref{unoccupied_kspace})
and (\ref{total_low}) together with the GA result \cite{rajdeep} for $n(\bk)$. We find
\be
\sum_{\bk}\int_{0}^{\Omega_L} d\omega A_{inc}(\bk,\omega) = {2x(1-x) \over U(1+x)}\sum_\bk \epsilon_{\bk} v^2_{\bk}
\label{incoherent_unocc_ksum}
\ee  

We have also obtained sum rule constraints on the density of states contributions coming separately
from the coherent and incoherent parts of the spectral function, however we omit these here.
In any case, in a $\bk$-integrated probe like tunneling it would be very hard to separate out the
coherent contribution from the incoherent one in experiments, unlike in ARPES where it seems possible
to do so. 

\medskip
\noindent{\bf Implications for Experiments}
\medskip

We first discuss the implications of our results for tunneling spectroscopy. 
The tunneling conductance in STM experiments is proportional to the local density of states 
$G(\br;eV) = {\rm constant} \times N(\br;\omega = eV)$ \cite{footnote2} where the
constant of proportionality involves tunneling matrix elements.
Thus our result (\ref{unoccupied_STM}) shows that the (energy integrated) positive bias conductance 
is small, of order $x$, while (\ref{occupied_STM}) implies that the (integrated) negative bias conductance 
is large, of order unity. This provides a qualitative explanation for the large asymmetry seen
in STM experiments which show a superconducting gap structure superimposed on a sloping ``background''
which decreases going from negative to positive bias. Our results predict how this asymmetry should
grow with underdoping (decreasing $x$). This asymmetry is most strikingly seen in the highly
underdoped {\em non}-superconducting cuprates such as 
Na$_x$Ca$_{1-x}$CuO$_2$Cl$_2$ studied by Hanaguri {\it et al.} \cite{davis1}. 
The nature of the ``zero temperature pseudogap state'' in such materials is an unsolved problem,
and in this context it is very important to re-emphasize that our results (\ref{occupied_STM}) and 
(\ref{unoccupied_STM}) make no assumptions about the broken symmetry in the ground state or the nature
of low-lying excitations.
 
In order to get quantitative information from STM experiments we 
look at ratios in which the unknown tunneling matrix elements cancel out.
Taking the ratio of the total unoccupied low-energy spectral weight (\ref{unoccupied_STM}) to the 
total occupied spectral weight (\ref{occupied_STM}) we obtain
\begin{equation}
{{\int_0^{\Omega_L} d\omega G(\br;\omega)} \over {\int_{-\infty}^0 d\omega G(\br;\omega)}} 
= {2x(\br) \over [1 - x(\br)]} + {{2\left\vert \langle K(\br) \rangle \right\vert} \over {U [1 - x(\br)]}}.
\label{ratio}
\end{equation}
The left hand side can now be estimated from STM data, provided one can
make a reasonable choice of the positive and negative high energy cutoffs
\cite{cutoffs}, and then used to infer the local hole doping $x(\br)$ from the
first term on the right hand side of (\ref{ratio}). The second term of order
$(xt/U)$ gives an estimate of the approximately 10 \% error made in estimating $x$.
 
In the second part of the paper we derived results for the doping dependence of the
coherent and incoherent parts of the spectral function. The predicted $x$ dependence 
of the coherent weight $Z$ of eq.~(\ref{z_factor}) has already been observed in
ARPES studies of nodal QP's (see ref.~\cite{paramekanti}). The important prediction
for the very small incoherent spectral weight on the unoccupied side 
(\ref{incoherent_unocc},\ref{incoherent_unocc_ksum})
should be testable in future inverse photoemission experiments once their energy resolution
is improved.

\medskip
{\bf Acknowledgments}
We have greatly benefited from many stimulating conversations with P. W. Anderson, who 
first got us interested in this problem.  We acknowledge helpful discussions with J. C. Davis
and thank him for keeping us abreast of his ongoing experiments. We also thank J. C. Campuzano, 
O. Fischer, L. H. Greene, P. Hentges, P. A. Lee and T. M. Rice for their valuable input.


\begin{thebibliography}{99}
\bibitem{cutoffs}
In analyzing experimental data, the cutoff $\Omega_L$ should be set by the charge transfer gap
$\approx 1.5 - 2.0$ eV. On the negative bias side, the ARPES dispersion in Bi2212 shows 
that the ``bottom of the band'' is $\approx - 0.5$ eV, but this is only the centroid of
a very broad spectral function. Other states begin to show up below about $- 1$ eV in 
ARPES \cite{shen,campuzano}. Thus replacing the ``$-\infty$'' by a 
cutoff $\approx - 1.0$ eV seems reasonable.
\end{thebibliography}
\end{document}